\documentclass[a4paper, oneside, twocolumn, notitlepage, 10pt]{extarticle_ecoc}
\usepackage{ecoc}
\usepackage[helvet]{sfmath}
\usepackage[per-mode=symbol]{siunitx}
\DeclareSIUnit{\sample}{Sa}
\DeclareSIUnit{\baud}{Bd}
\DeclareSIUnit{\bit}{bit}
\DeclareSIUnit{\fourd}{4D}
\DeclareSIUnit{\eightd}{8D}
\DeclareSIUnit{\dBm}{dBm}
\DeclareSIUnit{\dB}{dB}
\DeclareSIUnit{\bps}{bps}
\usepackage{glossaries}
\glsdisablehyper
\newcommand{\SetCapsType}{normalcaps}
\usepackage{xspace}  
\usepackage[shortcuts]{extdash}
\usepackage{listofitems,pgffor}    
\usepackage{siunitx}
\usepackage{xstring}    
\usepackage{silence}
\usepackage{xparse}

\setsepchar{;}  

\ifdefined\silencecommonwarnings
\else
	\def\silencecommonwarnings{true} 
\fi

\ifbool{\silencecommonwarnings}{%
    \WarningFilter{ECOtools}{Cannot define: DH}%
    \WarningFilter{ECOtools}{Cannot define: PAM}%
    \WarningFilter{ECOtools}{Cannot define: QAM}%
    \WarningFilter{ECOtools}{Cannot define: SI}%
    \WarningFilter{ECOtools}{Cannot define: PV}%
    \WarningFilter{ECOtools}{Cannot define: LP}%
    \WarningFilter{ECOtools}{Cannot define: uLP}%
    \WarningFilter{ECOtools}{Redefining DH}%
    }{}

\makeatletter
\providecommand{\SetCapsType}{smallcaps}

\long\def\@scTrue{smallcaps}
\long\def\@scFalse{normalcaps}
\newcommand{\acroSCaps}[1]{%
    \ifx\SetCapsType\@scTrue 
        \textsc{#1}%
    \else
        \MakeUppercase{#1}%
    \fi
}
\makeatother

\usepackage{scalerel}
\makeatletter
\newcommand\scslash{%
\ifx\SetCapsType\@scTrue 
    \protect\stretchrel*{$/$}{\textsc{e}}
\else
    /
\fi
} 
\makeatother 

\makeatletter
\@ifpackageloaded{babel}{%
    \newcommand{\usuk}[2]{%
        \iflanguage{USenglish}{#1}{#2}%
    }%
}{%
    \newcommand{\usuk}[2]{%
        #1%
    }%
}%

\newcommand{\langcheck}[2]{
    \@ifpackageloaded{babel}{%
        \iflanguage{USenglish}{#1}{#2}%
    }{%
        #1%
    }%
}

\makeatother

\newcommand{\short}[1]{%
    \glsentrytext{#1}\xspace%
}
\newcommand{\Short}[1]{%
    \Glsentrytext{#1}\xspace%
}
\newcommand{\normal}[1]{%
    \gls{#1}\xspace%
}
\newcommand{\longacr}[1]{%
    \acrlong{#1}\xspace%
}
\newcommand{\plural}[1]{%
    \glspl{#1}\xspace%
}
\newcommand{\full}[1]{%
    \acrfull{#1}\xspace%
}
\newcommand{\fullplural}[1]{%
    \acrfullpl{#1}\xspace%
}
\newcommand{\Normal}[1]{%
    \Gls{#1}\xspace%
}
\newcommand{\Plural}[1]{%
    \Glspl{#1}\xspace%
}
\newcommand{\Full}[1]{%
    \Acrfull{#1}\xspace%
}
\newcommand{\Fullplural}[1]{%
    \Acrfullpl{#1}\xspace%
} 

\newcommand{\texpdfif}[2]{%
    \ifcsname texorpdfstring\endcsname%
        \texorpdfstring{#1{#2}}{#2\xspace}%
    \else%
        #1{#2}%
    \fi%
}

\newcommand{\checkanddefine}[3]{%
	\ifcsname #1\endcsname%
        \PackageWarning{ECOtools}{Cannot define: #1 already defined, trying to define g#1 instead.}%
        \ifcsname g#1\endcsname%
            \PackageWarning{ECOtools}{Cannot define: g#1 also already defined.}%
    	\else%
        	\expandafter\newcommand\csname g#1\endcsname{%
        	    \texpdfif{#2}{#3}%
    	    }%
        \fi%
	\else%
    	\expandafter\newcommand\csname #1\endcsname{%
    	    \texpdfif{#2}{#3}%
	    }%
    \fi%
}

\newcommand{\redefine}[3]{%
    \PackageWarning{ECOtools}{Redefining #1}%
	\expandafter\renewcommand\csname #1\endcsname{%
	    \texpdfif{#2}{#3}%
    }%
}

\newcommand{\nAcronym}[4][]{%
	\newacronym[#1]{#2}{#3}{#4}%
	\checkanddefine{s#2}{\short}{#2}%
	\checkanddefine{#2}{\normal}{#2}%
	\checkanddefine{l#2}{\longacr}{#2}%
	\checkanddefine{#2s}{\plural}{#2}%
	\checkanddefine{f#2}{\full}{#2}%
	\checkanddefine{f#2s}{\fullplural}{#2}%
	\checkanddefine{su#2}{\Short}{#2}%
	\checkanddefine{u#2}{\Normal}{#2}%
	\checkanddefine{u#2s}{\Plural}{#2}%
	\checkanddefine{fu#2}{\Full}{#2}%
	\checkanddefine{fu#2s}{\Fullplural}{#2}%
	\IfStrEq{#2}{DH}{
	    \redefine{#2}{\normal}{#2}%
	    }{}%
}%

\NewDocumentCommand\qam{g}{%
    \IfNoValueTF{#1}{%
        \texpdfif{\gls}{QAM}\xspace%
        }{%
        \StrLen{#1}[\stringlength]%
        \ifnum\stringlength=0%
            \texpdfif{\gls}{QAM}\xspace%
        \else%
            {\qamlisthelper{#1}}%
        \fi%
        }%
}

\let\QAM\qam

\DeclareRobustCommand\qamlisthelper[1]{%
    \readlist*\args{#1}%
    \acroSCaps{\args[1]\=/}%
    \ifnum\argslen = 2%
        { and \acroSCaps{\args[2]}\=/}%
    \fi%
    \ifnum\argslen > 2%
        \foreach \n in {2,...,\argslen}{%
            \ifnum\n = \argslen%
                {, and }%
            \else 
                {, }%
            \fi%
            {\acroSCaps{\args[\n]}\=/}%
        }%
    \fi%
    \ifglsused{QAM}%
        {}%
        {ary }%
    \texpdfif{\gls}{QAM}%
}%

\NewDocumentCommand\pam{g}{%
    \IfNoValueTF{#1}{%
        \texpdfif{\gls}{PAM}\xspace%
        }{%
        \StrLen{#1}[\stringlength]%
        \ifnum\stringlength=0%
            \texpdfif{\gls}{PAM}\xspace%
        \else%
            {\pamlisthelper{#1}}%
        \fi%
        }%
}

\DeclareRobustCommand\pamlisthelper[1]{%
    \readlist*\args{#1}%
    \ifglsused{PAM}{%
        \texpdfif{\gls}{PAM}%
        \acroSCaps{\=/\args[1]}%
        \ifnum\argslen = 2%
            { and \=/\acroSCaps{\args[2]}}%
        \fi%
        \ifnum\argslen > 2%
            \foreach \n in {2,...,\argslen}{%
                \ifnum\n = \argslen%
                    {, and }%
                \else%
                    {, }%
                \fi%
                {\=/\acroSCaps{\args[\n]}}%
            }%
        \fi%
    }{%
        \acroSCaps{\args[1]\=/}%
        \ifnum\argslen = 2%
            { and \acroSCaps{\args[2]}\=/}%
        \fi%
        \ifnum\argslen > 2%
            \foreach \n in {2,...,\argslen}{%
                \ifnum\n = \argslen%
                    {, and }%
                \else%
                    {, }%
                \fi
                {\acroSCaps{\args[\n]}\=/}%
            }%
        \fi%
        {ary }%
        \texpdfif{\gls}{PAM}%
    }%
}%

\NewDocumentCommand\lp{g}{%
    \IfNoValueTF{#1}{%
        \texpdfif{\normal}{LP}%
        }{%
        \StrLen{#1}[\stringlength]%
        \ifnum\stringlength=0%
            \texpdfif{\normal}{LP}%
        \else%
            \ifglsused{LP}{}{\texpdfif{\normal}{LP}\xspace}%
            \lplisthelper[lp]{#1}%
        \fi%
        }%
}
\let\LP\lp%

\NewDocumentCommand\ulp{g}{%
    \IfNoValueTF{#1}{%
        \texpdfif{\Normal}{LP}\xspace%
        }{%
        \StrLen{#1}[\stringlength]%
        \ifnum\stringlength=0%
            \texpdfif{\Normal}{LP}\xspace%
        \else%
            \ifglsused{LP}{%
                \lplisthelper[Lp]{#1}%
            }{%
                \texpdfif{\Normal}{LP}\xspace\lplisthelper[lp]{#1}%
            }%
        \fi%
        }%
}
\DeclareRobustCommand\lplisthelper[2][lp]{%
    \readlist*\args{#2}%
    \foreach \n in {1,...,\argslen}{%
        \ifnum \n > 1%
            \ifnum \argslen > 2%
                {, }%
            \else%
                { }%
            \fi%
        \fi%
        \ifnum \n = \argslen%
            \ifnum \argslen > 1%
                {and }%
            \fi%
        \fi%
        \ifnum \n = 1%
            {\acroSCaps{#1}}
        \else%
            {\acroSCaps{\MakeLowercase{#1}}}%
        \fi%
        {\textsubscript{\StrSplit{\args[\n]}{2}{\csA}{\csB}\acroSCaps{\csA}\csB}}
    }%
}%

\nAcronym{128SPQAM}{\acroSCaps{128-sp-16-qam}}{128-ary set-partitioning \QAM{16}}

\nAcronym{2A8PSK}{\acroSCaps{2a8psk}}{2-ary amplitude 8-ary phaseshift keying}

\nAcronym{3CCMCF}{\acroSCaps{3cc-mcf}}{3-core coupled-core multi-core fiber}

\nAcronym{4D}{\acroSCaps{4d}}{four-dimensional}
\nAcronym{4D64PRS}{\acroSCaps{4d-64prs}}{\usuk{four-dimensional 64-ary polarization-ring-switching}{four-dimensional 64-ary polarisation-ring-switching}}
\nAcronym{4DOS128}{\acroSCaps{4d-os128}}{four-dimensional orthant-symmetric 128-ary modulation format}

\nAcronym{5B4D2A8PSK}{\acroSCaps{5b4d-2a8psk}}{5-bit four-dimensional two-amplitude 8-ary phase-shift keying}

\nAcronym{6B4D2A8PSK}{\acroSCaps{6b4d-2a8psk}}{6-bit four-dimensional two-amplitude 8-ary phase-shift keying}

\nAcronym{7B4D2A8PSK}{\acroSCaps{7b4d-2a8psk}}{7-bit four-dimensional two-amplitude 8-ary phase-shift keying}

\nAcronym{8D}{\acroSCaps{8d}}{eight-dimensional}\nAcronym{8D2048PRS}{\acroSCaps{8d-2048prs}}{eight-dimensional 2048-ary polarization-ring-switching}
\nAcronym{8D2048PRST1}{\acroSCaps{8d-2048prs-t1}}{eight-dimensional 2048-ary polarization-ring-switching type 1}
\nAcronym{8D2048PRST2}{\acroSCaps{8d-2048prs-t2}}{eight-dimensional 2048-ary polarization-ring-switching type 2}
\nAcronym{8DAPSK}{\acroSCaps{8d-apsk}}{eight-dimensional amplitude-phase-shift keying}

\nAcronym{ABC}{\acroSCaps{abc}}{automatic bias controller}
\nAcronym{AC}{\acroSCaps{ac}}{alternating current}
\nAcronym{ADC}{\acroSCaps{adc}}{analog-to-digital converter}
\nAcronym{AIR}{\acroSCaps{air}}{achievable information rate}
\nAcronym{AO}{\acroSCaps{ao}}{adaptive optics}
\nAcronym{AOM}{\acroSCaps{aom}}{acoustic optical modulator}
\nAcronym{APD}{\acroSCaps{apd}}{avalanche photodiode}
\nAcronym{API}{\acroSCaps{api}}{application programming interface}
\nAcronym{AR}{\acroSCaps{ar}}{achievable rate}
\nAcronym{ARRWG}{\acroSCaps{a}rr\acroSCaps{wg}}{arrayed-waveguide grating}
\nAcronym{ASE}{\acroSCaps{ase}}{amplified spontaneous emission}
\nAcronym{ASK}{\acroSCaps{ask}}{amplitude-shift keying}
\nAcronym{ASIC}{\acroSCaps{asic}}{application-specific integrated circuit}
\nAcronym{ATS}{\acroSCaps{ats}}{alignment tracking sensor}
\nAcronym{AWG}{\acroSCaps{awg}}{arbitrary-waveform generator}
\nAcronym{AWGN}{\acroSCaps{awgn}}{additive white Gaussian noise}

\nAcronym{BBU}{\acroSCaps{bbu}}{baseband unit}
\nAcronym{BCH}{\acroSCaps{bch}}{Bose-Chaudhuri-Hocquenghem}
\nAcronym{BER}{\acroSCaps{ber}}{bit error rate}
\nAcronym{BERT}{\acroSCaps{bert}}{bit error rate tester}
\nAcronym{BICM}{\acroSCaps{bicm}}{bit-interleaved coded modulation}
\nAcronym{BMD}{\acroSCaps{bmd}}{bit-metric decoding}
\nAcronym{BPD}{\acroSCaps{bpd}}{balanced photo-diode}
\nAcronym{BPF}{\acroSCaps{bpf}}{bandpass filter}
\nAcronym{BPS}{\acroSCaps{bps}}{blind phase search}
\nAcronym{BPSK}{\acroSCaps{bpsk}}{binary phase-shift keying}
\nAcronym{BRGC}{\acroSCaps{brgc}}{binary reflected Gray code}
\nAcronym{BTB}{\acroSCaps{btb}}{back-to-back}

\nAcronym{CAGR}{\acroSCaps{cagr}}{compound annual growth rate}
\nAcronym{CCDM}{\acroSCaps{ccdm}}{constant composition distribution matching}
\langcheck{%
    \nAcronym{CCF}{\acroSCaps{ccf}}{coupled-core fiber}%
    }{%
    \nAcronym{CCF}{\acroSCaps{ccf}}{coupled-core fibre}%
}%
\nAcronym{CD}{\acroSCaps{cd}}{chromatic dispersion}
\nAcronym{CIR}{\acroSCaps{cir}}{channel impulse response}
\nAcronym{CMA}{\acroSCaps{cma}}{constant modulus algorithm}
\nAcronym{CMF}{\acroSCaps{cmf}}{core multiplicity factor}
\nAcronym{CMUX}{\acroSCaps{cmux}}{core multiplexer}
\nAcronym{COTS}{\acroSCaps{cots}}{commercial off-the-shelf}
\nAcronym{ChUT}{\acroSCaps{chut}}{channel under test}
\nAcronym[firstplural=channels under test (\acroSCaps{cut}s)]{CUT}{\acroSCaps{cut}}{channel under test}
\nAcronym{CPE}{\acroSCaps{cpe}}{carrier phase estimation}
\nAcronym{CPU}{\acroSCaps{cpu}}{central processing unit}
\nAcronym{CSPR}{\acroSCaps{cspr}}{carrier-to-signal power ratio}
\nAcronym{CUDA}{\acroSCaps{cuda}}{compute unified device architecture}
\nAcronym{CW}{\acroSCaps{cw}}{continuous wave}
\nAcronym{CCD}{\acroSCaps{ccd}}{charge-coupled device}

\nAcronym{DA}{\acroSCaps{da}}{driver amplifier}
\nAcronym{DAC}{\acroSCaps{dac}}{digital-to-analog converter}
\nAcronym{DC}{\acroSCaps{dc}}{direct current}
\nAcronym{DBP}{\acroSCaps{dbp}}{digital backpropagation}
\nAcronym{DCF}{\acroSCaps{dcf}}{\usuk{dispersion-compensating fiber}{dispersion-compensating fibre}}
\langcheck{%
    \nAcronym{DCI}{\acroSCaps{dci}}{data center interconnect}
    }{%
    \nAcronym{DCI}{\acroSCaps{dci}}{data centre interconnect}
}%
\nAcronym{DDLMS}{\acroSCaps{dd-lms}}{decision-directed least mean square}
\nAcronym{DEMUX}{\acroSCaps{demux}}{de-multiplexer}
\nAcronym{DFA}{\acroSCaps{dfa}}{\usuk{doped fiber amplifier}{doped fibre amplifier}}
\nAcronym{DFB}{\acroSCaps{dfb}}{distributed feedback}
\nAcronym{DGD}{\acroSCaps{dgd}}{differential group delay}
\nAcronym{DH}{\acroSCaps{dh}}{digital holography}
\nAcronym{DM}{\acroSCaps{dm}}{distribution matcher}
\nAcronym{DMR}{\acroSCaps{dm}}{dichroic mirror}
\nAcronym{DMA}{\acroSCaps{dma}}{direct memory access}
\nAcronym{DMD}{\acroSCaps{dmd}}{differential mode delay}
\nAcronym{DMG}{\acroSCaps{dmg}}{differential modal gain}
\nAcronym{DMGD}{\acroSCaps{dmgd}}{differential mode group delay}
\nAcronym{DML}{\acroSCaps{dml}}{directly-modulated laser}
\nAcronym{DP}{\acroSCaps{dp}}{\usuk{dual-polarization}{dual-polarisation}}
\nAcronym{DPC}{\acroSCaps{dpc}}{digital pre-compensation}
\nAcronym{DPE}{\acroSCaps{dpe}}{digital pre-emphasis}
\nAcronym{DPIQ}{\acroSCaps{dp-iqm}}{\usuk{dual-polarization \acroSCaps{iq}-modulator}{dual-polarisation \acroSCaps{iq}-modulator}}
\nAcronym{DPLL}{\acroSCaps{dpll}}{digital phase-locked loop}
\nAcronym{DQPSK}{\acroSCaps{dqpsk}}{differential quaternary phase-shift-keying}
\nAcronym{DRA}{\acroSCaps{dra}}{distributed Raman amplifier}
\nAcronym{DRE}{\acroSCaps{dre}}{digital resolution enhancer}
\nAcronym{DSB}{\acroSCaps{dsb}}{double-sideband}
\nAcronym{DSF}{\acroSCaps{dsf}}{\usuk{dispersion-shifted fiber}{dispersion-shifted fibre}} 
\nAcronym{DSO}{\acroSCaps{dso}}{digital sampling oscilloscope}
\nAcronym{DSP}{\acroSCaps{dsp}}{digital signal processing}
\nAcronym{DUT}{\acroSCaps{dut}}{device-under-test}
\nAcronym{DWDM}{\acroSCaps{dwdm}}{dense wavelength-division multiplexing}

\nAcronym{EAM}{\acroSCaps{eam}}{electro-absorption modulator}
\nAcronym{ECL}{\acroSCaps{ecl}}{external cavity laser}
\nAcronym{ED}{\acroSCaps{ed}}{Eucledian distance}
\nAcronym{EDF}{\acroSCaps{edf}}{\usuk{erbium-doped fiber}{erbium-doped fibre}}
\nAcronym{EDFA}{\acroSCaps{edfa}}{\usuk{erbium-doped fiber amplifier}{erbium-doped fibre amplifier}}
\nAcronym{ENOB}{\acroSCaps{enob}}{effective number of bits}
\nAcronym{ESS}{\acroSCaps{ess}}{enumerative sphere shaping}

\langcheck{%
    \nAcronym{FBG}{\acroSCaps{fbg}}{fiber Bragg grating}%
    }{%
    \nAcronym{FBG}{\acroSCaps{fbg}}{fibre Bragg grating}%
}%
\nAcronym{FD}{\acroSCaps{fd}}{frequency domain}
\nAcronym{FDE}{\acroSCaps{fde}}{\usuk{frequency domain equalizer}{frequency domain equaliser}}
\nAcronym{FEC}{\acroSCaps{fec}}{forward error correction}
\nAcronym{FFE}{\acroSCaps{ffe}}{\usuk{feed-forward equalizer}{feed-forward equaliser}}
\nAcronym{FFT}{\acroSCaps{fft}}{fast Fourier transform}
\nAcronym{FIR}{\acroSCaps{fir}}{finite impulse response}
\nAcronym{FLOPS}{\acroSCaps{flops}}{floating point operations per second}
\nAcronym{FMEDF}{\acroSCaps{fm-edf}}{\usuk{few-mode erbium-doped fiber}{few-mode erbium-doped fibre}}
\nAcronym{FMEDFA}{\acroSCaps{fm-edfa}}{\usuk{few-mode erbium-doped fiber amplifier}{few-mode erbium-doped fibre amplifier}}
\langcheck{%
    \nAcronym{FMF}{\acroSCaps{fmf}}{few-mode fiber}%
    }{%
    \nAcronym{FMF}{\acroSCaps{fmf}}{few-mode fibre}%
}%
\nAcronym[plural=FM-MCF, firstplural=\usuk{few-mode multi-core fibers}{few-mode multi-core fibres}]{FMMCF}{\acroSCaps{fm-mcf}}{\usuk{few-mode multi-core fiber}{few-mode multi-core fibre}}
\langcheck{%
    \nAcronym{FMPBGF}{\acroSCaps{fm-pbgf}}{few-mode photonic bandgap fiber}%
    }{%
    \nAcronym{FMPBGF}{\acroSCaps{fm-pbgf}}{few-mode photonic bandgap fibre}%
}%
\nAcronym{FOV}{\acroSCaps{fov}}{field of view}
\nAcronym{FPGA}{\acroSCaps{fpga}}{field-programmable gate array}
\nAcronym{FWM}{\acroSCaps{fwm}}{four-wave mixing}
\nAcronym{FSO}{\acroSCaps{fso}}{free-space optics}
\nAcronym{FUT}{\acroSCaps{fut}}{\usuk{fiber under test}{fibre under test}}

\nAcronym{GD}{\acroSCaps{gd}}{group delay}
\nAcronym{GI}{\acroSCaps{gi}}{graded-index}
\nAcronym{GFF}{\acroSCaps{gff}}{gain flattening filter}
\nAcronym{GIFMF}{\acroSCaps{gi-fmf}}{\usuk{graded-index few-mode fiber}{graded-index few-mode fibre}}
\nAcronym{GIMMF}{\acroSCaps{gi-mmf}}{\usuk{graded-index multi-mode fiber}{graded-index multi-mode fibre}}
\nAcronym{GMI}{\acroSCaps{gmi}}{\usuk{generalized mutual information}{generalised mutual information}}
\nAcronym{GNSE}{\acroSCaps{gnse}}{generalized nonlinear Schr\"{o}dinger equation}
\nAcronym{GPU}{\acroSCaps{gpu}}{graphics processing unit}
\nAcronym{GS}{\acroSCaps{gs}}{geometric shaping}
\nAcronym{GV}{\acroSCaps{gv}}{group velocity}
\nAcronym{GVD}{\acroSCaps{gvd}}{group velocity dispersion}
\nAcronym{GPIO}{\acroSCaps{gpio}}{general purpose input output}
\nAcronym{GUI}{\acroSCaps{gui}}{graphical user interface}

\langcheck{%
    \nAcronym{HCF}{\acroSCaps{hcf}}{hollow-core fiber}
    }{%
    \nAcronym{HCF}{\acroSCaps{hcf}}{hollow-core fibre}
}%
\nAcronym{HDFEC}{\acroSCaps{hd-fec}}{hard-decision forward error correction}
\nAcronym{HG}{\acroSCaps{hg}}{Hermite-Gaussian}
\nAcronym{HV}{\acroSCaps{hv}}{Hufnagel-Valley}
\nAcronym{HAP}{\acroSCaps{hap}}{Hufnagel-Andrew-Phillips}

\nAcronym{ICS}{\acroSCaps{ics}}{inter-core skew}
\nAcronym{ICXT}{\acroSCaps{ic-xt}}{inter-core cross-talk}
\nAcronym[plural=IL, firstplural=insertion losses (\acroSCaps{il})]{IL}{\acroSCaps{il}}{insertion loss}
\nAcronym{IFFT}{\acroSCaps{ifft}}{inverse fast Fourier transform}
\nAcronym{IMDD}{\acroSCaps{im}\scslash \acroSCaps{dd}}{intensity-modulation direct-detection}
\nAcronym{IQM}{\acroSCaps{iqm}}{in-phase and quadrature modulator}
\nAcronym{ISI}{\acroSCaps{isi}}{inter-symbol interference}

\nAcronym{JGN}{\acroSCaps{jgn}}{Japan Gigabit Network}

\nAcronym{KK}{\acroSCaps{kk}}{Kramers-Kronig}
\nAcronym{KIT}{\acroSCaps{kit}}{Karlsruhe Institute of Technology}

\nAcronym{LCOS}{\acroSCaps{LCoS}}{liquid crystal on silicon}
\nAcronym{LDPC}{\acroSCaps{ldpc}}{low-density parity-check}
\nAcronym{LEAF}{\acroSCaps{leaf}}{\usuk{large effective area fiber}{large effective area fibre}}
\nAcronym{LFSR}{\acroSCaps{lfsr}}{linear-feedback shift register}
\nAcronym{LG}{\acroSCaps{lg}}{Laguerre-Gaussian}
\nAcronym{LMS}{\acroSCaps{lms}}{least means squares}
\nAcronym{LLR}{\acroSCaps{llr}}{log-likelihood ratio}
\nAcronym{LO}{\acroSCaps{lo}}{local oscillator}
\nAcronym{LP}{\acroSCaps{lp}}{\usuk{linearly polarized}{linearly polarised}}
\nAcronym{LSPS}{\acroSCaps{lsps}}{\usuk{loop-synchronized polarization scrambler}{loop-synchronised polarisation scrambler}}
\nAcronym{LUT}{\acroSCaps{lut}}{lookup table}

\nAcronym{MB}{\acroSCaps{mb}}{Maxwell-Bolzmann}
\langcheck{%
    \nAcronym{MCF}{\acroSCaps{mcf}}{multi-core fiber}%
    }{%
    \nAcronym{MCF}{\acroSCaps{mcf}}{multi-core fibre}%
}%
\nAcronym{MDG}{\acroSCaps{mdg}}{mode dependent gain}
\nAcronym[firstplural=mode-dependent losses (\acroSCaps{mdl})]{MDL}{\acroSCaps{mdl}}{mode-dependent loss}
\nAcronym{MDM}{\acroSCaps{mdm}}{mode-division multiplexing}
\nAcronym{MEMS}{\acroSCaps{mems}}{micro-electro-mechanical systems}
\nAcronym{MF}{\acroSCaps{mf}}{matched filter}
\nAcronym{MFD}{\acroSCaps{mfd}}{mode field diameter}
\nAcronym{MI}{\acroSCaps{mi}}{mutual information}
\nAcronym{MIMO}{\acroSCaps{mimo}}{multiple-input multiple-output}
\nAcronym{ML}{\acroSCaps{ml}}{machine learning}
\nAcronym{MMA}{\acroSCaps{mma}}{multi-modulus algorithm}
\nAcronym{MMEDF}{\acroSCaps{mmedf}}{\usuk{multi-mode erbium-doped fiber}{multi-mode erbium-doped fibre}}
\nAcronym{MMEDFA}{\acroSCaps{mmedfa}}{\usuk{multi-mode erbium-doped fiber amplifier}{multi-mode erbium-doped fibre amplifier}}
\nAcronym{MMF}{\acroSCaps{mmf}}{\usuk{multi-mode fiber}{multi-mode fibre}}
\nAcronym{MMSE}{\acroSCaps{mmse}}{minimum mean squared error}
\nAcronym{MP}{\acroSCaps{mp}}{minimum phase}
\nAcronym{MPLC}{\acroSCaps{mplc}}{multi-plane light converter}
\nAcronym{MRC}{\acroSCaps{mrc}}{maximum ratio combining}
\nAcronym{MSE}{\acroSCaps{mse}}{mean squared error}
\nAcronym{MUX}{\acroSCaps{mux}}{multiplexer}
\nAcronym{MZM}{\acroSCaps{mzm}}{Mach-Zehnder modulator}
\nAcronym{MZI}{\acroSCaps{mzi}}{Mach-Zehnder interferometer}

\nAcronym{NA}{\acroSCaps{na}}{numerical aperture}
\langcheck{%
    \nAcronym{NANF}{\acroSCaps{nanf}}{nested antiresonant nodeless fiber}%
    }{%
    \nAcronym{NANF}{\acroSCaps{nanf}}{nested antiresonant nodeless fibre}%
}%
\nAcronym{NF}{\acroSCaps{nf}}{noise figure}
\nAcronym{NGMI}{\acroSCaps{ngmi}}{\usuk{normalized generalized mutual information}{normalised generalised mutual information}}
\nAcronym{NLSE}{\acroSCaps{nlse}}{nonlinear Schr\"{o}ding equation}
\nAcronym{NN}{\acroSCaps{nn}}{neural network}
\nAcronym{NIC}{\acroSCaps{nic}}{network interface card}
\nAcronym{NICT}{\acroSCaps{nict}}{National Institute of Information and Communications Technology}
\nAcronym{NIR}{\acroSCaps{nir}}{near-infrared}
\nAcronym{NRZ}{\acroSCaps{nrz}}{non-return-to-zero}
\nAcronym{NZDSF}{\acroSCaps{nz-dsf}}{\usuk{non-zero dispersion-shifted fiber}{non-zero dispersion-shifted fibre}} 

\nAcronym{OAM}{\acroSCaps{oam}}{orbital angular momentum}
\nAcronym{OBTB}{\acroSCaps{obtb}}{optical back-to-back}
\nAcronym{OCT}{\acroSCaps{oct}}{outer cladding thickness}
\nAcronym{ODE}{\acroSCaps{ode}}{ordinary differential equation}
\nAcronym{ODL}{\acroSCaps{odl}}{optical delay line}
\nAcronym{OEO}{\acroSCaps{oeo}}{optical-electrical-optical}
\nAcronym{OFC}{\acroSCaps{ofc}}{Optical Fiber Communications Conference}
\nAcronym{OFDR}{\acroSCaps{ofdr}}{optical frequency-domain reflectometer} 
\nAcronym{OFDM}{\acroSCaps{ofdm}}{orthogonal frequency division multiplexing}
\nAcronym{OH}{\acroSCaps{oh}}{overhead}
\nAcronym{OMFT}{\acroSCaps{omft}}{optical multi-format transmitter}
\nAcronym{OOK}{\acroSCaps{ook}}{on-off keying}
\nAcronym{OPLL}{\acroSCaps{opll}}{optical phase-locked loop}
\nAcronym{OSA}{\acroSCaps{osa}}{\usuk{optical spectrum analyzer}{optical spectrum analyser}}
\nAcronym{OSNR}{\acroSCaps{osnr}}{optical signal-to-noise ratio}
\nAcronym{OTDR}{\acroSCaps{otdr}}{optical time-domain reflectometer}
\nAcronym{OTF}{\acroSCaps{otf}}{optical tunable filter}
\langcheck{%
    \nAcronym{OVNA}{\acroSCaps{ovna}}{optical vector network analyzer}%
    }{%
    \nAcronym{OVNA}{\acroSCaps{ovna}}{optical vector network analyser}%
}%
\nAcronym{OTG}{\acroSCaps{otg}}{optical turbulence generator}

\nAcronym{PAM}{\acroSCaps{pam}}{pulse-amplitude modulation}
\nAcronym{PAS}{\acroSCaps{pas}}{probabilistic amplitude shaping}
\nAcronym{PAPR}{\acroSCaps{papr}}{peak-to-average power ratio}
\nAcronym{PBC}{\acroSCaps{pbc}}{\usuk{polarization beam combiner}{polarisation beam combiner}}
\langcheck{%
    \nAcronym{PBGF}{\acroSCaps{pbgf}}{photonic bandgap fiber}%
    }{%
    \nAcronym{PBGF}{\acroSCaps{pbgf}}{photonic bandgap fibre}%
}%
\nAcronym{PBS}{\acroSCaps{pbs}}{polarization beam splitter}
\nAcronym{PC}{\acroSCaps{pc}}{physical contact}
\nAcronym{PCVD}{\acroSCaps{pcvd}}{plasma chemical vapor depostion}
\nAcronym{PCG}{\acroSCaps{pcg64}}{64-bit permuted congruential generator}
\nAcronym{PD}{\acroSCaps{pd}}{photodiode}
\langcheck{%
    \nAcronym{PDL}{\acroSCaps{pdl}}{polarization-dependent loss}
    }{%
    \nAcronym{PDL}{\acroSCaps{pdl}}{polarisation-dependent loss}
}%
\nAcronym{PDM}{\acroSCaps{pdm}}{\usuk{polarization-division multiplexing}{polarisation-division multiplexing}}
\langcheck{%
    \nAcronym{PER}{\acroSCaps{per}}{polarization extinction ratio}%
    }{%
    \nAcronym{PER}{\acroSCaps{per}}{polarisation extinction ratio}%
}%
\nAcronym{PIC}{\acroSCaps{pic}}{photonic integrated circuit}
\nAcronym{PL}{\acroSCaps{pl}}{photonic lantern}
\nAcronym{PMBPSK}{\acroSCaps{pm-bpsk}}{polarization-multiplexed binary phase-shift keying}
\nAcronym{PMQPSK}{\acroSCaps{pm-qpsk}}{polarization-multiplexed quaternary phase-shift keying}
\nAcronym{PM8QAM}{\acroSCaps{pm-8qam}}{polarization-multiplexed 8-ary quadrature amplitude modulation}
\nAcronym{PMD}{\acroSCaps{pmd}}{\usuk{polarization mode dispersion}{polarisation mode dispersion}}
\langcheck{%
    \nAcronym{PMF}{\acroSCaps{pmf}}{polarization-maintaining fiber}%
    }{%
    \nAcronym{PMF}{\acroSCaps{pmf}}{polarization-maintaining fibre}%
}%
\nAcronym{PMP}{\acroSCaps{pmp}}{phase-matching point}
\nAcronym{PNOB}{\acroSCaps{pnob}}{physical number of bits}
\nAcronym{PON}{\acroSCaps{pon}}{passive-optical network}
\nAcronym{PRBS}{\acroSCaps{prbs}}{pseudorandom bit sequence}
\nAcronym{PROFA}{\acroSCaps{profa}}{\usuk{pitch reducing optical fiber array}{pitch reducing optical fibre array}}
\nAcronym{PPM}{\acroSCaps{ppm}}{pulse-position modulation}
\nAcronym{PS}{\acroSCaps{ps}}{probabilistic shaping}
\langcheck{%
    \nAcronym{PSCF}{\acroSCaps{pscf}}{pure silica core fiber}%
    }{%
    \nAcronym{PSCF}{\acroSCaps{pscf}}{pure silica core fibre}%
}%
\nAcronym{PSD}{\acroSCaps{psd}}{power spectral density}
\nAcronym{PSF}{\acroSCaps{psf}}{point spread function}
\nAcronym{PSK}{\acroSCaps{psk}}{phase-shift keying}
\nAcronym{PSP}{\acroSCaps{psp}}{\usuk{principal states of polarization}{principal states of polarisation}}
\nAcronym{PSW}{\acroSCaps{psw}}{\usuk{polarization switch}{polarisation switch}}
\nAcronym{PID}{\acroSCaps{pid}}{proportional–integral–derivative}
\nAcronym{PV}{\acroSCaps{pv}}{process value}

\nAcronym{QAM}{\acroSCaps{qam}}{quadrature amplitude modulation}
\nAcronym{QPSK}{\acroSCaps{qpsk}}{quaternary phase-shift keying}
\nAcronym{QSM}{\acroSCaps{qsm}}{quasi-single-mode}

\nAcronym{RAM}{\acroSCaps{ram}}{random-access memory}
\nAcronym{RC}{\acroSCaps{rc}}{raised cosine}
\nAcronym{RCMF}{\acroSCaps{rcmf}}{relative core multiplicity factor}
\nAcronym{RF}{\acroSCaps{rf}}{radio frequency}
\nAcronym{RI}{\acroSCaps{ri}}{refractive index}
\nAcronym{RLS}{\acroSCaps{rls}}{recursive least squares}
\nAcronym{RRC}{\acroSCaps{rrc}}{root-raised-cosine}
\nAcronym{ROADM}{\acroSCaps{roadm}}{reconfigurable optical add-drop multiplexer}
\nAcronym{ROI}{\acroSCaps{roi}}{region of interest} 
\nAcronym{RZDBPSK}{\acroSCaps{rz-dbpsk}}{return-to-zero differential binary phase-shift keying} 
\nAcronym{RZDQPSK}{\acroSCaps{rz-dqpsk}}{return-to-zero differential quaternary phase-shift keying} 

\nAcronym{S2}{\acroSCaps{S\textsuperscript{2}}}{spatially and spectrally resolved}
\nAcronym{SA}{\acroSCaps{sa}}{simulated annealing}
\nAcronym{SamPerSym}{\acroSCaps{sps}}{samples per symbol}
\nAcronym{SBS}{\acroSCaps{sbs}}{stimulated Brillouin scattering}
\nAcronym{SCM}{\acroSCaps{scm}}{subcarrier multiplexing}
\nAcronym{SDFEC}{\acroSCaps{sd-fec}}{soft-decision forward error correction}
\nAcronym{SDM}{\acroSCaps{sdm}}{space-division multiplexing}
\nAcronym{SE}{\acroSCaps{se}}{spectral efficiency}
\nAcronym{SER}{\acroSCaps{ser}}{symbol error rate}
\nAcronym{SI}{\acroSCaps{si}}{step index}
\nAcronym{SIFMF}{\acroSCaps{si-fmf}}{\usuk{step-index few-mode fiber}{step-index few-mode fibre}}
\nAcronym{SISMF}{\acroSCaps{si-smf}}{\usuk{step-index single-mode fiber}{step-index single-mode fibre}}
\nAcronym{SLM}{\acroSCaps{slm}}{spatial light modulator}
\nAcronym{SMF}{\acroSCaps{smf}}{\usuk{single-mode fiber}{single-mode fibre}}
\nAcronym{SMUX}{\acroSCaps{smux}}{spatial multiplexer}
\nAcronym{SNR}{\acroSCaps{snr}}{signal-to-noise ratio}
\nAcronym{SOA}{\acroSCaps{soa}}{semiconductor optical amplifier}
\nAcronym[firstplural=\usuk{states of polarization (\acroSCaps{sop})}{states of polarisation (\acroSCaps{sop})}]{SOP}{\acroSCaps{sop}}{\usuk{state of polarization}{state of polarisation}}
\nAcronym{SPM}{\acroSCaps{spm}}{self-phase modulation}
\nAcronym{SPS}{\acroSCaps{sps}}{samples per symbol}
\nAcronym{SRS}{\acroSCaps{srs}}{stimulated Raman scattering}
\nAcronym{SSB}{\acroSCaps{ssb}}{single-sideband}
\nAcronym{SSBI}{\acroSCaps{ssbi}}{signal-signal beat interference}
\nAcronym{SSFM}{\acroSCaps{ssfm}}{split-step Fourier method}
\nAcronym{SSMF}{\acroSCaps{ssmf}}{\usuk{standard single-mode fiber}{standard single-mode fibre}}
\nAcronym{STAXT}{\acroSCaps{staxt}}{short-term average cross-talk}
\nAcronym{STL}{\acroSCaps{stl}}{swept tunable laser}
\nAcronym{SVD}{\acroSCaps{svd}}{singular value decomposition}
\nAcronym{SW}{\acroSCaps{sw}}{sequence-wise}
\nAcronym{SWI}{\acroSCaps{swi}}{swept wavelength interferometry}
\nAcronym{SP}{\acroSCaps{sp}}{setpoint}

\nAcronym{TD}{\acroSCaps{td}}{time domain}
\nAcronym{TDE}{\acroSCaps{tde}}{\usuk{time domain equalizer}{time domain equaliser}}
\nAcronym{TDM}{\acroSCaps{tdm}}{time-domain multiplexing}
\nAcronym{TDMSDM}{\acroSCaps{tdm-sdm}}{time-domain multiplexed space-division multiplexing}
\nAcronym{TE}{\acroSCaps{te}}{transverse electric}
\nAcronym{TEC}{\acroSCaps{tec}}{thermally-expanded-core}
\nAcronym{TIA}{\acroSCaps{tia}}{trans-impedance amplifier}
\nAcronym{TM}{\acroSCaps{tm}}{transverse magnetic}
\nAcronym{TH4D}{\acroSCaps{th-4d}}{time domain hybrid four-dimensional}
\nAcronym{TH4D2A8PSK}{\acroSCaps{th-4d-2a8psk}}{time-domain hybrid four-dimensional two-amplitude eight-phase-shift keying}

\nAcronym{VCSEL}{\acroSCaps{vcsel}}{vertical-cavity surface emitting laser}
\nAcronym{VHDL}{\acroSCaps{vhdl}}{\acroSCaps{Vhsic} Hardware Description Language}
\nAcronym{VOA}{\acroSCaps{voa}}{variable optical attenuator}

\nAcronym{WDM}{\acroSCaps{wdm}}{wavelength-division multiplexing}
\nAcronym{WFS}{\acroSCaps{wfs}}{wavefront sensor}
\nAcronym{WGA}{\acroSCaps{wga}}{weakly guiding approximation}
\nAcronym{WGN}{\acroSCaps{wgn}}{white Gaussian noise}
\nAcronym[firstplural=wavelength selective switches (\acroSCaps{wss}s)]{WSS}{\acroSCaps{wss}}{wavelength selective switch}
\nAcronym{WC}{\acroSCaps{wc}}{weakly coupled}
\nAcronym{WCMCF}{\acroSCaps{wcmcf}}{\usuk{weakly-coupled multi-core fiber}{weakly-coupled multi-core fibre}}

\nAcronym{XPM}{\acroSCaps{xpm}}{cross-phase modulation}
\langcheck{%
    \nAcronym{XPOLM}{\acroSCaps{xp}ol\acroSCaps{m}}{cross-polarization modulation}
    }{%
    \nAcronym{XPOLM}{\acroSCaps{xp}ol\acroSCaps{m}}{cross-polarisation modulation}
}%
\nAcronym{XT}{\acroSCaps{xt}}{cross-talk}

\nAcronym{3DWG}{\acroSCaps{3dwg}}{3D-waveguide} 

\usepackage{subcaption}

\usepackage[capitalise]{cleveref}
\usepackage[export]{adjustbox}

\begin{document}
\selectlanguage{english}    


\title{Experimental Investigation of Mode Diversity Reception Using an Optical Turbulence Generator and Digital Holography
}%


\author{Vincent~van~Vliet\textsuperscript{(*)},
        Menno~van~den~Hout,
        Sjoerd~van~der~Heide,
        and Chigo~Okonkwo
}

\maketitle                  



    \begin{strip}
        \begin{author_descr}
        
            High-Capacity Optical Transmission Laboratory, Electro-Optical Communications Group,\\
            Eindhoven University of Technology, the Netherlands
            \textsuperscript{(*)}{\,\uline{v.v.vliet@student.tue.nl}}\\
        \end{author_descr}
    \end{strip}

\setstretch{1.1}
\renewcommand\footnotemark{}
\renewcommand\footnoterule{}


\begin{strip}
  \begin{ecoc_abstract}
 {Mode diversity reception is experimentally investigated using an optical turbulence generator, off-axis digital holography, and digital demultiplexing. The results confirm improved fibre coupling efficiency when receiving the optical field using a multi-mode fibre instead of a single-mode fibre under turbulent conditions, specifically beam wander. The coupling loss is reduced by receiving additional modes. \textcopyright\,\,2022~The~Author(s)}
  \end{ecoc_abstract}
\end{strip}


\section{Introduction}
Free-space optical (FSO)\glsunset{FSO} communications is considered a promising alternative to \lRF based communications because of its improved bandwidth, capacity, security against eavesdropping, equipment size, weight, and power consumption. Applications which can benefit from this include backhaul for wireless networks, last mile access, and space communications \cite{Kaushal2017}.

The usage of high-quality, commercially available fibre-optic components can catalyse the development of \FSO communication systems. However, the plethora of these components is \SMF-based. It has been shown that free-space-to-\SMF coupling severely suffers from turbulent conditions \cite{Dikmelik:05}. The resulting signal fades cause degradation of the link quality. Recently, theoretical models \cite{fan2021, Fardoost:19}, simulations \cite{Zheng:16, Krimmer:2020}, and experiments \cite{Zheng:16, Fontaine2019, Arikawa:20, Geisler2016, Huang2021} have demonstrated the potential of mode diversity reception as a technique to improve the fibre coupling efficiency. Here, additional spatial modes are received using a \MMF, after which a \lDEMUX and coherent recombination provide compatibility with \SMF-based components. As turbulence induces higher-order aberrations, accommodation of higher-order modes improves the coupling efficiency as compared to single mode. Before-mentioned experiments were performed using either static phase screens or uncharacterised turbulence. 

In this work, we evaluate the performance of mode diversity reception using naturally evolving turbulence produced in an \OTG, an off-axis \DH measurement setup, and digital demultiplexing. The main induced turbulent effect is beam wander. It is shown that the coupling loss statistics are improved when receiving additional modes as compared to only the fundamental mode. The coupling loss is reduced by \SI{1.15}{dB} for an outage probability of~\SI{0.1}{\percent} by receiving the first 2 mode groups instead of only the fundamental mode.

\begin{figure}
   \centering
        \includegraphics[width=0.99\linewidth]{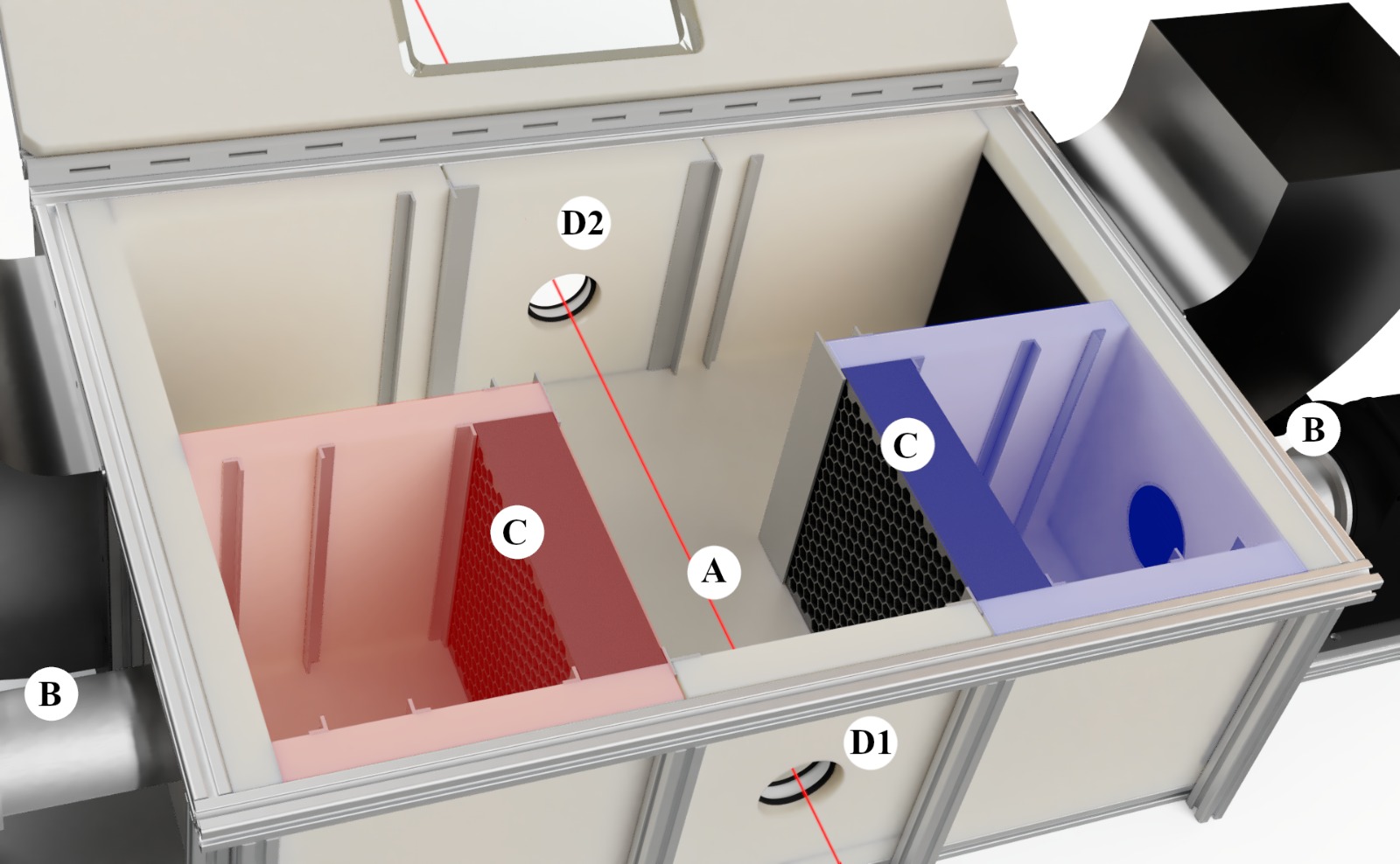}
    \caption{Design of the \OTG used in this work \cite{OTGthesis}. (A): mixing chamber, (B): (hot-)air blowers, (C): metal mesh to laminarise flow, (D1) and (D2): laser window panels.}
    \label{fig:OTG_render}
\end{figure}


\section{Experimental setup}
\begin{figure*}
   \centering
    \includegraphics[width=0.8\linewidth]{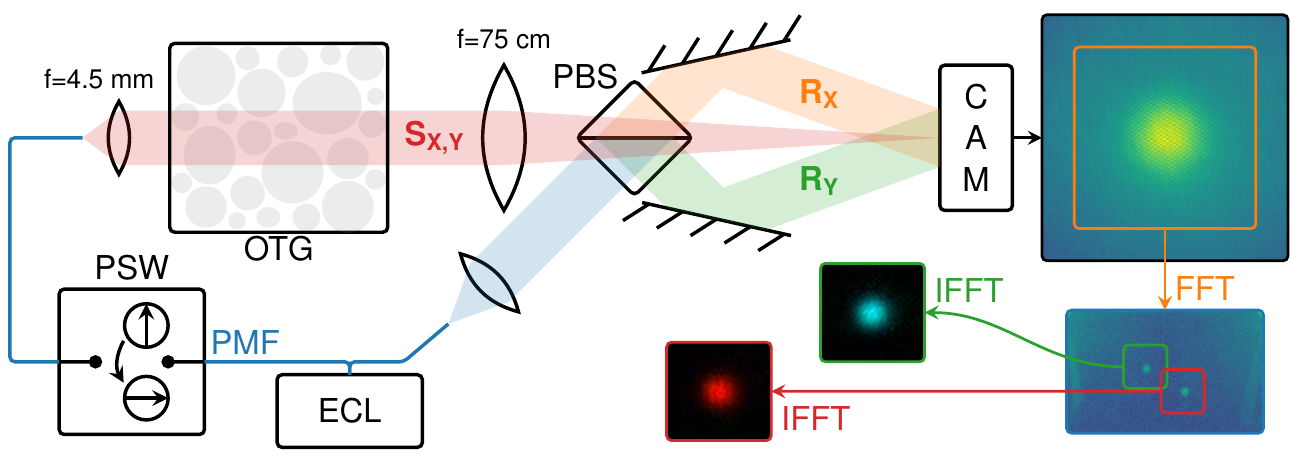}
    \includegraphics[width=0.17\linewidth]{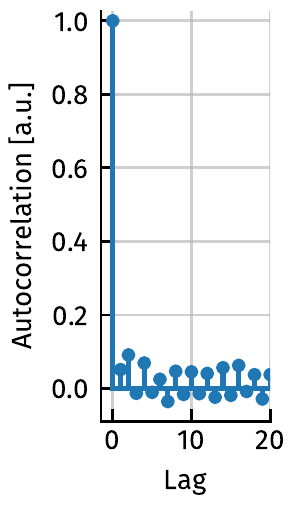}
    \caption{\textbf{(left)} Experimental setup employing the \OTG and optical off-axis \DH setup and digital field extraction. S\textsubscript{X,Y}: signal distorted by the \OTG, R\textsubscript{X} and R\textsubscript{Y}: reference beams, ECL: external cavity laser, PMF: polarisation-maintaining fibre, PBS: polarisation beam splitter, (I)FFT: (inverse) fast Fourier transform. Note that S\textsubscript{X,Y} passes over the PBS. \textbf{(right)} Autocorrelation of beam centroid displacement in $x$-direction.
    }
    \label{fig:setup}
\end{figure*}

The experimental free-space optical setup is shown in \cref{fig:setup}~(left). A signal beam \textbf{S\textsubscript{X\&Y}} is propagated through the \OTG and captured with a camera operating at \SI{333}{fps}. The signal is alternated between $x$- and $y$-polarisation every frame using a \PSW. \textbf{S\textsubscript{X\&Y}} has a diameter of \SI{0.88}{mm} after collimation. Using an off-axis \DH measurement setup \cite{demoOFC,thesisSjoerdDH}, the complete optical field including amplitude and phase is measured for both polarisations. To this end, \textbf{S\textsubscript{X\&Y}} is interfered with \textbf{R\textsubscript{X,Y}}, two orthogonally-polarised reference beams with a flat wavefront. Subsequent \DSP allows for the extraction of the $x$- and $y$-polarised complex fields from the interference pattern.

\textbf{S\textsubscript{X\&Y}} is distorted by turbulence in the \OTG. This is a device in which turbulence is produced by mixing two air flows in a compact chamber. The temperature difference between the two flows is used to control the strength of the produced turbulence. The design of the \OTG is shown in \cref{fig:OTG_render}. Using the beam wander variance characterisation technique \cite{Kaushal2011ExperimentalConditions}, the device is characterised for a ground-to-satellite link scenario. More information on the \OTG and the characterisation process can be found in \cite{OTGthesis}.

The fibre coupling is performed in the digital domain by demultiplexing the measured optical fields into target mode fields of the fibre of interest, here we use a \SI{50}{\um} core diameter \GI \MMF \cite{Sillard50um}. The target modes are calculated using a numerical scalar \LP mode solver. The measured optical fields are digitally scaled such that the unperturbed transmitted field couples optimally  to the \LP{01} mode of the considered fibre. This is the digital equivalent of a beam expander to maximize the coupling efficiency between the \LP{01} mode of the transmitter \SMF to the \LP{01} mode of the considered receiver \MMF.

It is assumed that every captured frame represents an independent instance of the channel. The autocorrelation function of the beam centroid displacement in the $x$-direction is shown on the right of \cref{fig:setup}. Although the frames are not completely uncorrelated, the rapid decrease towards 0 makes our assumption valid.


\section{Results}
\begin{figure}[h]
    \captionsetup[subfigure]{labelformat=empty}
    \centering
    \subfloat[\label{fig:histograms_a}]
    {
        \includegraphics[width=0.99\linewidth]{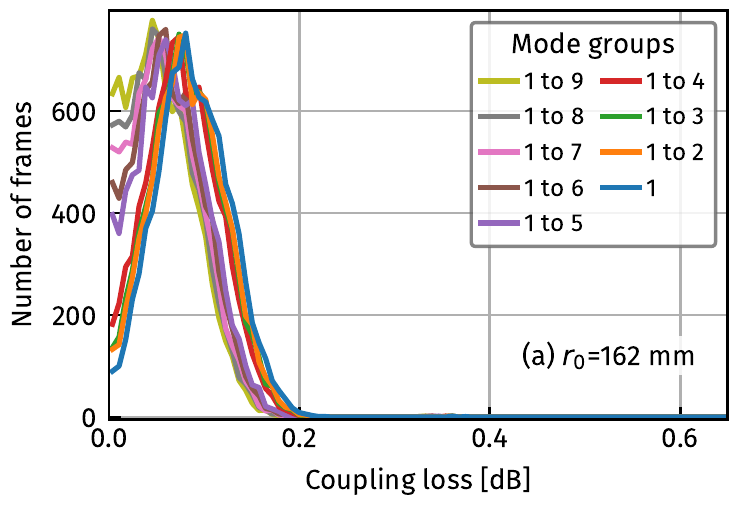}
    }\\
    \subfloat[\label{fig:histograms_b}]
    {
        \includegraphics[width=0.99\linewidth]{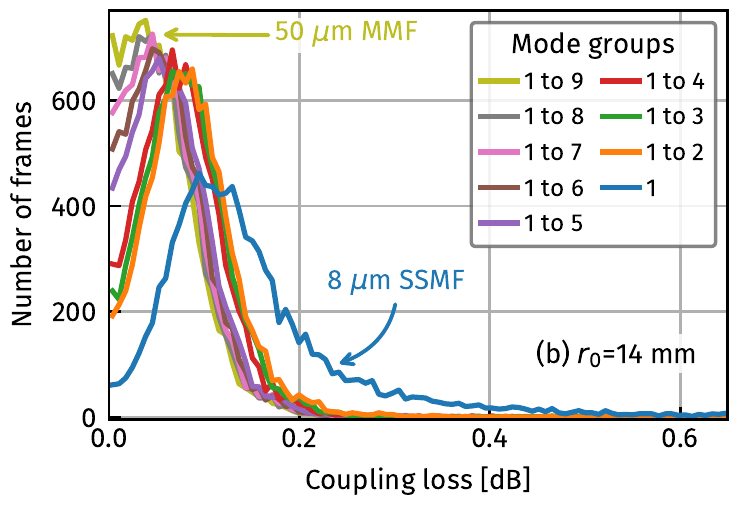}
    }
    \caption{Histogram of coupling loss over 10,000 frames for\\(a) $r_0=\SI{162}{mm}$ and (b) $r_0=\SI{14}{mm}$}
    \label{fig:histograms}
\end{figure}

The complex fields of the distorted signal were measured during 30~seconds for various turbulence strengths, resulting in \SI{10000}{frames} for each measurement. \cref{fig:histograms} shows the histogram of the resulting coupling losses after the digital demultiplexing procedure for turbulence conditions with $r_0~=~\SI{162}{mm}$ (\cref{fig:histograms_a}) and $r_0~=~\SI{14}{mm}$ (\cref{fig:histograms_b}), the weakest and strongest turbulence produced in the \OTG, respectively. Here, $r_0$ is the turbulence coherence length, also known as the Fried parameter\cite{Fried:66}, which is a measure of turbulence strength in an optical path. It can be seen that for weak turbulence conditions the coupling performance of a \SMF closely resembles that of a \MMF. However, coupling to the \SMF worsens after propagation through stronger turbulence. In that case, the coupling efficiency improves when the receiving fibre supports a higher number of modes.

\begin{figure}
    \captionsetup[subfigure]{labelformat=empty}
    \centering
    \subfloat[\label{fig:histograms_outage_a}]
    {
        \includegraphics[width=0.99\linewidth]{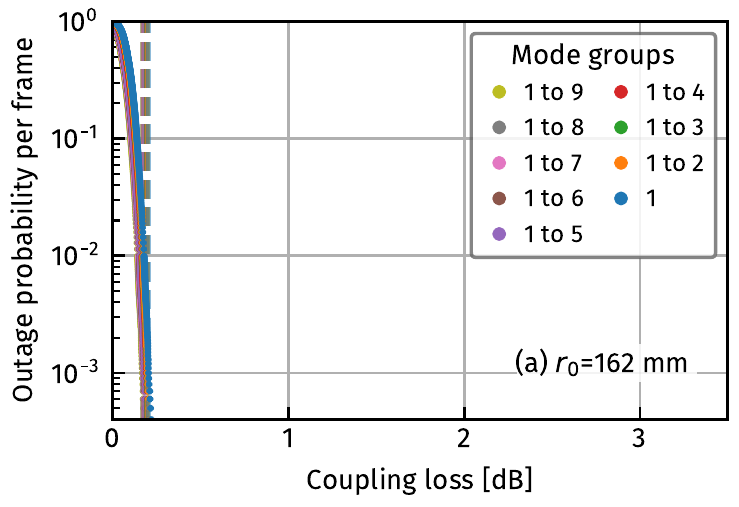}
    }\\
    \subfloat[\label{fig:histograms_outate_b}]
    {
        \includegraphics[width=0.99\linewidth]{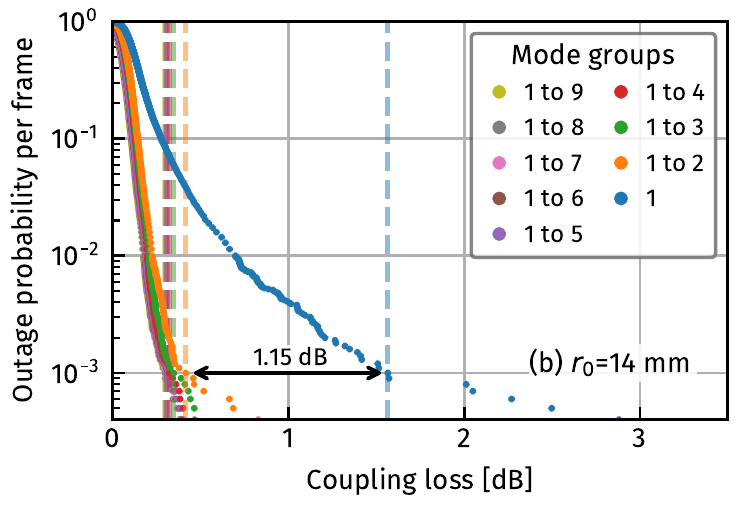}
    }
    \caption{Outage probability statistics from 10000 frames for (a) $r_0=\SI{162}{mm}$ and (b) $r_0=\SI{14}{mm}$. The vertical lines indicate the threshold loss corresponding to an outage probability of \SI{0.1}{\percent}.}
    \label{fig:histograms_outage}
\end{figure}

\cref{fig:histograms_outage} displays the corresponding outage probabilities. No performance difference can be distinguished for the weakest turbulence condition (\cref{fig:histograms_outage_a}). However, a clear distinction is observed for the stronger turbulence condition (\cref{fig:histograms_outate_b}). In that case, for an outage probability of \SI{0.1}{\percent} the coupling loss is reduced by \SI{1.15}{dB} when coupling into a fibre supporting three modes, as compared to the \SMF.

Because the beam diameter of \textbf{S\textsubscript{X\&Y}} is much smaller than $r_0$, the main effect induced by the turbulence is beam wander. This turbulent effect is represented by the \LP{11a} and \LP{11b} modes. Accordingly, the results show that incorporation of the second mode group clearly improves the coupling efficiency. However, consideration of the higher-order modes does not provide significant advantage because of the lack of higher-order aberrations in the optical field. It is expected that receiving higher-order modes is advantageous when these aberrations are present.

\section{Conclusions}
The coupling loss statistics of free-space-to-fibre coupling are investigated experimentally using an optical turbulence generator, off-axis digital holography, and digital demultiplexing. It was shown that mode diversity reception improves the coupling efficiency after propagation through turbulent atmosphere. Considering three modes instead of only the fundamental mode results in a reduction in coupling loss of \SI{1.15}{dB} for an outage probability per frame of \SI{0.1}{\percent}.

Future steps include the characterisation of the \OTG turbulence for satellite-to-ground and horizontal \FSO links and extension to a deployed field link for long-term testing in a realistic scenario. Furthermore, measurements with an increased beam diameter will provide information on the fibre coupling efficiency for fields with turbulence-induced higher-order aberrations.


\section{Acknowledgements}

Funding received from KPN-TU/e Smart Two Program, the Dutch NWO Gravitation Program (Grant Number 024.002.033) and NWO Perspectief project Optical Wireless Superhighways: Free photons.


\printbibliography

\vspace{-4mm}

\end{document}
